# Inner relaxations in equiatomic single-phase high-entropy Cantor alloy

Alevtina Smekhova[1,*], Alexei Kuzmin[2], Konrad Siemensmeyer[1], Radu Abrudan[1], Uwe Reinholz[3], Ana Guilherme Buzanich[3], Mike Schneider[4], Guillaume Laplanche[4], Kirill V. Yusenko[3,*]

[1] *Helmholtz-Zentrum Berlin für Materialien und Energie (HZB), D-12489 Berlin, Germany*

[2] *Institute of Solid State Physics, University of Latvia, LV-1063 Riga, Latvia*

[3] *Bundesanstalt für Materialforschung und – prüfung (BAM), D-12489 Berlin, Germany*

[4] *Institut für Werkstoffe, Ruhr-Universität Bochum, D-44801 Bochum, Germany*

**Abstract:**

The superior properties of high-entropy multi-functional materials are strongly connected with their atomic heterogeneity through many different local atomic interactions. The detailed element-specific studies on a local scale can provide insight into the primary arrangements of atoms in multicomponent systems and benefit to unravel the role of individual components in certain macroscopic properties of complex compounds. Herein, multi-edge X-ray absorption spectroscopy combined with reverse Monte Carlo simulations was used to explore a homogeneity of the local crystallographic ordering and specific structure relaxations of each constituent in the equiatomic single-phase face-centered cubic CrMnFeCoNi high-entropy alloy at room temperature. Within the considered fitting approach, all five elements of the alloy were found to be distributed at the nodes of the fcc lattice without any signatures of the additional phases at the atomic scale and exhibit very close statistically averaged interatomic distances (2.54 – 2.55 Å) with their nearest-neighbors. Enlarged structural displacements were found solely for Cr atoms. The macroscopic magnetic properties probed by conventional magnetometry demonstrate no opening of the hysteresis loops at 5 K and illustrate a complex character of the long-range magnetic order after field- assisted cooling in ± 5 T. The observed magnetic behavior is assigned to effects related to structural relaxations of Cr. Besides, the advantages and limitations of the reverse Monte Carlo approach to studies of multicomponent systems like high-entropy alloys are highlighted.

Graphical abstract

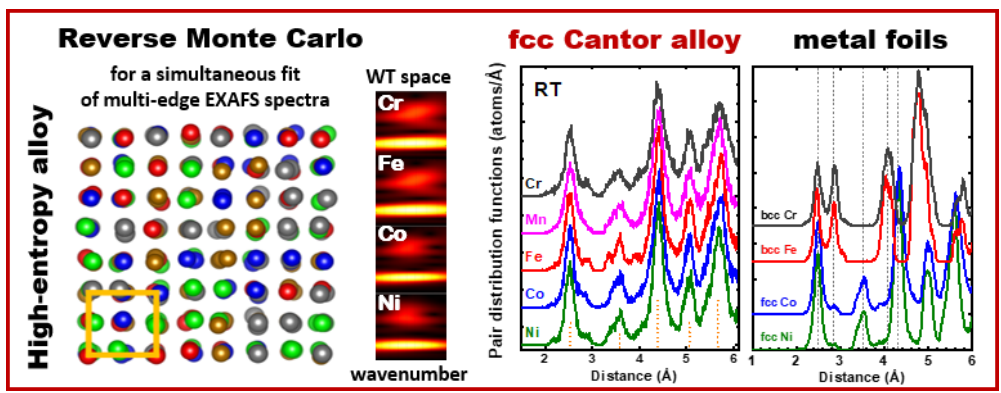

**Keywords:** high-entropy alloys, reverse Monte Carlo (RMC), element-specific spectroscopy, extended X-ray absorption fine structure (EXAFS), X-ray absorption near edge structure (XANES), magnetism

## 1. Introduction

High-entropy and compositionally complex alloys based on 3d-block elements are recognized to be promising functional materials that exhibit high potential for different aspects in materials science [1-3]. Nowadays, they are also established to be very attractive for practical applications in vital renewable energy technologies like energy storage and conversion [4]. High-entropy alloys (HEAs) are known for their high chemical complexity resulting in different local atomic environments [5], crystallographic structures and mechanisms of phase formation [6], smaller diffusion mobility of atoms [5, 7], and enhanced mechanical properties [8]. Their impressive performance in the fields of hydrogen storage [9, 10], noble-metal-free electrocatalysts [11, 12], oxygen evolution and reduction reactions [13, 14], carbon dioxide conversion [15], and supercapacitors with virtually unlimited lifecycle [16, 17], *etc* has been already uncovered. Besides, an enhanced radiation resistance [18] suggests the successful use of HEAs in advanced nuclear applications [19].

Depending on composition and processing routes, HEAs exhibit variations in their phase constitution and microscopic morphology [20]. Several HEA systems based on Cr, Mn, Fe, Co, Ni and Al were proposed to be single-phase face-centered cubic (fcc) structures in quite broad compositional and temperature intervals [21-23]. The equiatomic CrMnFeCoNi HEA – the classical Cantor alloy – forms a stable fcc solid solution above ~800 °C that can be retained down to room temperature (RT), if the alloy is cooled at a sufficiently high rate [24, 25], whereas $Al_x$CrFeCoNi HEAs for $x < 0.3$ are fcc at RT after cooling rapidly from above 1000 °C [26-30]. These features make these alloys suitable model HEAs for the investigation of various fundamental aspects related to short- and long-range ordering. Moreover, they allow investigating the relationships between local aggregations and local lattice distortions with macroscopic properties such as phase stability, hardness, ductility, corrosion, and magnetism.

The equiatomic single-phase fcc CrMnFeCoNi HEA shows high-temperature structural stability together with good strength and ductility at low temperatures [8, 31, 32]. High corrosion resistance due to the formation of a $Cr_2O_3$-rich passivation layer has been reported for alkali and acidic media [33, 34]. Magnetic properties of the CrMnFeCoNi HEA were recently reported in Ref. [35], where two magnetic transformations below 100 K (the paramagnetic to spin-glass and the ferromagnetic one) were found while the fcc structure was maintained down to 3 K. An extraordinary mechanical and chemical stability of CrMnFeCoNi HEA under high-temperature and high-pressure conditions were investigated in many details either experimentally or theoretically in Refs. [24, 36-39].

To improve the mechanical properties of the Cantor alloy while retaining its fcc structure, off-equiatomic compositions were investigated. For instance, Bracq et al. [40] studied the isopleths of each consistent element (e.g. $Mn_x(CrFeCoNi)_{100-x}$) to identify the compositional and temperature ranges in which fcc solid solutions are formed, and the mechanical properties of these alloys were systematically investigated [41-43]. It was found that these systems with a modified stoichiometry can form single-phase fcc solid solutions within a broad composition range as 0–25 at.% for Cr, 0–50 at.% for Mn, 0–50 at.% for Fe, 10–50 at.% for Co and 10–100 at.% for Ni. Nevertheless, the influence of each element on the stability and constitution of HEAs needs to be further addressed for the efficient design of new materials.

A large number of different local atomic configurations in HEAs results from the multitude of how the constituent elements are distributed around a particular atom within the several coordination shells. For the fcc lattice, these local arrangements within the first two coordination shells affect strongly the central atom. Indeed, different atomic sizes and different local atomic environments of atoms, in turn, lead to lattice distortions that are commonly considered as significant: in the original Cantor alloy, lattice distortions are expected to be on a scale of up to ~0.03 Å [44] whereas, for the modified Cantor alloys containing, for example, Al and Pd, such distortions could be larger [45, 46]. Therefore, a large number of different local atomic configurations in multicomponent single-phase fcc HEAs play an important role in properties that depend strongly on local atomic interactions [44]: among them, vacancy migration and diffusion, dislocation slip, and magnetic properties could be mentioned.

Local lattice distortions are responsible for the solid solution strengthening in HEAs since they can introduce energy barriers against dislocation motion due to random fluctuations [47-50]. Recently, lattice distortions in the Cantor alloy were studied by neutron diffraction at ambient conditions [51], by element-specific extended X-ray absorption fine structure (EXAFS) and by density functional theory (DFT) calculations [52-54]. In Ref. [51] the datasets were well described by considering only a single atom type with a compositionally weighted average scattering length of all the constituent elements (a grey atom model) while in Ref. [54], the main attention was paid already to element-resolved distortions of each constituent. It was found that, while distortions themselves are quite small on average (~0.1%), their fluctuations are an order of magnitude larger (up to 2–3%), especially for Cr and Mn. The DFT calculations were performed considering the magnetic ordering and were compared to the non-spin-polarized one to especially validate the impact of magnetism because the results of recent *ab initio* calculations [49] pointed out that in addition to the configurational entropy, other entropy contributions like electronic, vibrational, and magnetic excitations are of similar importance. It was concluded that magnetism is responsible for the unique character of local bond fluctuations and is critically required for realistic DFT simulations. In Ref. [53], the atomic pair distances between individual atoms of the alloy revealed that Mn atoms have a slightly larger bond distance (~0.4%) with their neighbors. An attempt to distinguish the local structural disorder from the thermal contribution was done by combining the low-temperature neutron diffraction and X-ray techniques. In Ref. [52] using temperature-dependent EXAFS, it was demonstrated that atomic size, charge transfer, magnetism, and local ordering are the important factors that affect the displacement of an individual atom in a multi-component alloy, i.e. element-specific structural relaxations. At the same time, an attempt to use neutron diffraction data and a series of atomistic simulations together with *ab initio* calculations [55] has shown that the short-range order in a four-component CrFeCoNi alloy is coupled with size mismatch of different constituents and lattice distortions. A brief overview of other methods used for lattice distortion studies in HEAs can be found in Refs. [56] and [57, 58]; the latter ones include also the theory background for quantitative comparison of lattice strains (static lattice distortions) revealed by different experimental techniques. It is important to notice that microscopic and X-ray diffraction studies performed for various HEA systems cannot reliably address the constituent arrangements at the atomic scale; thus, a demand for different element-specific studies at the absorption edges supported by a proper data evaluation suitable for the multicomponent systems is currently of increasing interest.

Recently, we have employed the element-specific EXAFS technique in conjunction with the reverse Monte Carlo (RMC) based analysis to reveal key peculiarities of individual atom distributions on the local scale up to several coordination shells in $Al_x$CrFeCoNi [59]. The interatomic distances between different nearest-neighbor pairs and quantitative estimations of local atomic structure relaxations were found. To avoid the well-known drawbacks of the conventional EXAFS analysis based on multi-shell fitting that could lead to unreliable and strongly biased results for multicomponent systems such as HEAs due to a large number of fitting parameters and strong correlation between them, a valuable advantage of atomistic simulations based on the RMC method was exploited. In particular, we used the possibility to fit simultaneously the same structural model to wavelet transforms of experimental EXAFS spectra taken at several K absorption edges of the constituents. In such a way, the pair distribution functions associated with local crystallographic configurations of each constituent element and local inhomogeneities attributed to a particular type of atoms were uncovered.

In the present work, the equiatomic Cantor alloy was chosen as an appropriate model for element-specific studies of inner structure relaxations to probe the details of local configurations at the atomic scale. The peculiarities of individual pair distribution functions have been revealed by the RMC method applied to EXAFS spectra taken at the Cr, Fe, Co, and Ni K absorption edges at RT. Conventional magnetometry was used to illustrate a very complex magnetic ordering formed inside the Cantor alloy highlighting the presence of the competing ferro- and antiferromagnetic interactions. Besides, the degree of surface atoms oxidation was probed by X-ray absorption spectroscopy at the $L_{2,3}$ absorption edges of all constituents.

## 2. Materials and Methods

### 2.1. Sample preparation and initial characterization

The Cantor alloy was prepared by vacuum induction melting using high purity metals ( ≥ 99.9 wt.%). It was then homogenized at 1200 °C for 48 h in an evacuated quartz tube under $3 \times 10^{-5}$ mbar and rotary swaged at RT. The obtained rods with a 17-mm diameter were recrystallized at 1020 °C for 1 h, resulting in an average grain size of ~50 µm and homogeneous chemical composition. More details about our preparation route can be found in Refs. [60, 61]. For further characterization, flakes were chipped off the alloy by mechanical milling to produce a powder-like sample. X-ray diffraction (XRD) was performed using this sample at RT and the data were collected in θ -2θ Bragg-Brentano geometry in the 2θ range of 20–90° on an Empyrean machine (Panalytical) using Cu K$_\alpha$ radiation ($\lambda$ = 1.54 Å) and the PIXcel 1D detector (Fig. 1). The profiles were refined by applying the LeBail model-free fitting routine. The XRD patterns from the CrMnFeCoNi alloy confirm its single-phase character with the fcc lattice (*Fm-3m* space group) and unit cell parameter *a* = 3.58214(3) Å. Its elemental composition (at.%) is Cr$_{20.4}$Mn$_{20.1}$Fe$_{19.6}$Co$_{19.7}$Ni$_{19.9}$ as determined by electron microprobe analysis.

**2.2 X-ray absorption spectroscopy at the K absorption edges**

X-ray absorption spectra of the HEA (Fig. 2a) were recorded by measuring fluorescence denoted as fluorescent yield (FY) at the Cr (5.99 keV), Fe (7.11 keV), Co (7.71 keV), and Ni (8.33 keV) K absorption edges at the BAM*line* of the Helmholtz-Zentrum Berlin (BESSY II, Berlin, Germany) [62]. The measurements were carried out with horizontally polarized hard X-rays from a 7 T wavelength shifter (WLS) at 21-22 °C in air. The samples in the form of 0.1-0.3 mm flakes were fixed in the X-ray beam between two 50-µm thick Kapton foils. A custom-made four-element energy-dispersive detector was used to collect the fluorescence signal in a similar way as in Ref. [63]. X-ray absorption spectra of pure metallic foils were recorded for reference by FY at the same absorption edges (Fig. 2b). Besides, the X-ray absorption spectrum of Mn foil was collected in transmission mode. The EXAFS spectra of pure metallic foils collected in transmission mode at the K edges of other constituents at the same beamline can be found in Ref. [59].

Before RMC analysis, the EXAFS spectra $\chi(k)k^2$ at the K edges of Cr, Mn, Fe, Co, and Ni were extracted from the raw data using a conventional procedure [64] implemented in the XAESA code [65]. For the considered HEA, the EXAFS spectra recorded at only four K absorption edges of 3d constituents (Cr, Fe, Co, Ni) were used in the RMC simulations due to the limited quality of the Mn K-edge EXAFS spectrum.

**2.3 Reverse Monte Carlo (RMC) analysis**

The analysis of EXAFS spectra of the fcc HEA collected at four K-edges of Cr, Fe, Co, and Ni was performed using the reverse Monte Carlo (RMC) method based on the evolutionary algorithm (EA) implemented in the EvAX code [66, 67]. Compared to conventional EXAFS analysis, which may bias results in the case of multicomponent systems, the RMC method is believed to be more robust because it allows the same structural model to be simultaneously fitted to several EXAFS spectra collected independently at different absorption edges, taking into account multiple scattering (MS) effects, as well as contributions from distant and overlapping coordination shells. At the same time, one should remember that EXAFS spectra are sensitive to the atomic distribution functions, so, only the relative positions of atoms obtained during the RMC simulation are important and not their absolute coordinates.

For the RMC simulations, a starting structural model (a simulation box) was constructed in a form of a supercell with a size of $4a \times 4a \times 4a$ ($a$ = 3.582 Å) employing the periodic boundary conditions (Fig. 3). The experimental value of the lattice parameter $a$ refined from the XRD data was used and the size of the supercell was kept constant during RMC simulations. The Cr, Mn, Fe, Co, and Ni atoms were randomly distributed in a proper concentration at the Wyckoff positions of the fcc supercell including 256 atoms (52 Cr, 52 Mn, 50 Fe, 51 Co, 51 Ni).

In the current study, we followed the same procedure as described previously in Ref. [59]. The number of atomic configurations simultaneously considered in the EA was 32, and the largest allowed displacement of atoms from their initial position was 0.4 Å, which is sufficient to describe both static and thermal disorder. The fixed supercell size and small value of the allowed atom displacements play the role of constraints and stabilize the structural model. As a result, the final model does not deviate much from the crystallographic structure but takes into account both thermal and static disorders present in the alloy.

At each RMC iteration, the configuration-averaged K-edge EXAFS spectra $\chi(k)k^2$ were calculated over all Cr, Fe, Co, and Ni atoms in the simulation box, and their Morlet wavelet transforms (WTs) were compared with those of the experimental EXAFS spectra. The best agreement between the Morlet WTs of the experimental and calculated EXAFS spectra was used as a criterion for the model structure optimization.

The WT calculations were performed in the $k$-space range from 3.0 to 10 Å$^{-1}$ (or 12 Å$^{-1}$) and in the $R$-space range from 1.0 to 6.0 Å. The number of RMC iterations was 5000 to guarantee the convergence of the structural model; no significant improvements in the residuals were observed beyond this number. Each RMC simulation resulted in a set of atomic coordinates, which was subsequently used to calculate the pair distribution functions (PDFs) $g(r)$, the mean interatomic distances $r$, the mean square relative displacements (MSRDs) for each pair of atoms, and the mean square displacements (MSDs) for atoms of each type. Calculated in such a way PDFs are element-specific projections of the three-dimensional structure of a material over a radial distance relative to the chosen absorbing atom, where only the amount and the type of atoms located at the particular distance from the absorber are important. So, there is no need to generate billions of local configurations expected for the five-component system, because they will very probably result in the same or very similar PDFs. Moreover, since only one structural model (the same starting configuration) is used in the RMC simulation during the simultaneous fit to four experimental EXAFS spectra collected at different absorption edges (these spectra act as the independent experimental data sets) and the fit is performed in the wavelet space, the small differences in the scattering amplitudes of 3d constituents are automatically taken into account, and the contributions from all principal components are properly accounting. Thus, the final structural model and the extracted PDFs represent a self-consistent solution that is in good agreement with all available experimental EXAFS data in both $k$ and $R$ spaces. To improve statistics, 12 sets of different (independent) starting structural models were considered for final PDFs, so the final PDFs describe the configuration-averaged local environment around a particular absorber in the most unbiased manner. Despite a total number of local configurations considered in the fit does not cover all possible relative arrangements of five different constituents, the fit outcomes like interatomic distances within the first coordination shell of absorbers and their mean square displacements can be still considered as statistically averaged quantities.

During RMC simulations the configuration-averaged EXAFS spectra were calculated for atoms of each type (all constituents) using *ab initio* real-space multiple-scattering FEFF8.50L code [68, 69] including the MS effects up to the 4$^{th}$ order. The scattering potential and partial phase shifts were calculated within the muffin-tin (MT) approximation [68, 69] for each absorption edge only once, considering the cluster with a radius of 4.8 Å centered at the required absorbing metal atom (Cr, Fe, Co, or Ni). Small changes in the cluster potential induced by atom displacements during the RMC/EA simulations were neglected. The photoelectron inelastic losses were accounted for within the one-plasmon approximation using the complex exchange- correlation Hedin-Lundqvist potential [70]. The amplitude reduction factor $S^2$ was included in the scattering amplitude calculated by the FEFF code [68, 69]; no additional correction of the EXAFS amplitude was performed. For each absorption edge, the values of the $E_0$ energies used in the definition of the photoelectron wavenumber $k = [(2m_e/\hbar^2)(E - E_0)]^{0.5}$ were set to those determined carefully in advance by performing the RMC simulations of reference compounds (pure metallic foils).

**2.4 Magnetometry**

Magnetometry measurements were carried out using a commercial magnetic properties measurement system (MPMS, Quantum Design) in magnetic fields up to 5 T at 5 K. The field-assisted cooling in ± 5 T was followed by the measurements of magnetic hysteresis loops without additional setting of the magnetic field to zero. The vibrating sample magnetometer (VSM) mode was used with a frequency of 14 Hz and 2 mm amplitude. The sample was firmly pressed to avoid any movements of flakes kept in a plastic capsule. The sample mass (22.33 mg) was measured with a microbalance.

**2.5 X-ray absorption spectroscopy at the $L_{2,3}$ absorption edges**

X-ray absorption near edge structure (XANES) spectra (see Fig. A1 in the Appendix) were collected at the Cr (560–640 eV), Mn (630–670 eV), Fe (690–740 eV), Co (770–820 eV), and Ni (840–890 eV) $L_{2,3}$ absorption edges in the total electron yield mode (TEY) at RT with linearly polarized soft X-rays. The HEA flakes were mounted into the ALICE chamber [71, 72] attached to the UE52_SGM undulator beamline [73] at the BESSY II synchrotron storage ring operated by Helmholtz-Zentrum Belin (HZB). The intensity of the incoming X-ray photons was monitored by the current from a gold-coated plane elliptical mirror upstream of the experimental chamber. The typical basic pressure inside the chamber during the experiment was in the range of $10^{-7}$-$10^{-8}$ mbar. The results of this complementary investigation can be found in the Appendix.

**3. Results and discussion**

**3.1 EXAFS spectroscopy**

To probe the peculiarities of the averaged local environments of 3d constituent elements in the Cantor alloy, the element-specific X-ray absorption spectroscopy was exploited at the K absorption edges of Cr, Fe, Co and Ni at RT. X-ray absorption spectra recorded from the HEA flakes as well as from the reference metallic foils are shown in Fig. 2 and represent the local coordination of a particular type of atoms in the bulk volume. The shape and positions of EXAFS oscillations found above each absorption edge for the considered HEA (Fig. 2a) indicate a dominantly homogeneous local environment of Cr, Fe, Co and Ni absorbers adopted in the fcc structure. Moreover, the spectra of all elements in the Cantor alloy are very similar to those of pure Co and Ni foils shown in Ref. [59].

In Fig. 2b, the intensity of EXAFS oscillations for the reference foils measured in the fluorescence mode is significantly suppressed due to self-absorption effects. Nevertheless, the frequencies of EXAFS oscillations are the same as compared with the spectra measured in transmission [59], and therefore a clear distinction between spectra recorded for the bcc (Cr, Fe) and fcc/hcp (Ni/Co) crystallographic structures could be made (see inset in Fig. 2b). The Mn K-edge EXAFS spectrum of Mn foil is significantly different from the others and corresponds to the cubic α-Mn phase (space group *I-43m* (217)) as it will be shown below.

A qualitative comparison of EXAFS spectra for reference foils with those of the HEA allows one to conclude that the majority of the elements adopt the fcc structure of the considered Cantor alloy independently of their crystallographic structure as pure metals.

**3.2 Reverse Monte Carlo simulations**

Due to a high heterogeneity at the atomic level, the number of local configurations around absorbers of each type in HEAs is extremely large (Fig. 3). However, the main peculiarities of atom distributions within the several first coordination shells of each absorber could be described as a statistical average by a set of pair distribution functions $g(r)$ found using the RMC approach. In the case of the fcc structure, the first peak located between 1.0 and 3.0 Å in the Fourier transforms (FTs) of the experimental EXAFS spectra (Fig. 4) originates predominantly from the contributions of the twelve atoms of the first coordination shell and is well isolated allowing a high accuracy of quantitative estimations related to the nearest-neighbors around absorbers of each type.

Moreover, the possibility to simultaneously fit the same structural model to several EXAFS spectra of different constituents offers an opportunity to overcome the limitations of the conventional multi-shell EXAFS analysis related to the Nyquist criterion [64], i.e. the large number of strongly correlated fitting parameters. It can be shown (e.g. Ref. [59]) that EXAFS spectra, recorded in the $k$-space range from 3.0 to 12 Å$^{-1}$ for the fcc system, exhibit the first peak in FT in the range mentioned above and would allow one to determine only ten parameters in the parametric EXAFS model. So, no more than three coordination shells could be effectively fitted considering the total number of nearest-neighbours $N$, radial distance to the nearest-neighbors $r$, and mean square relative displacement $\sigma^2$ of the absorber as free parameters. The RMC method allows us to fit a larger number of shells including different multi-scattering processes with high accuracy even if the EXAFS data for one of the constituent elements are missing (as Mn in the current case).

An important difference and advantage of the RMC fit from other atomistic simulations as, for example, molecular dynamics or DFT calculations is that the constructed structural model with the relaxed positions of all atoms represents the experimental EXAFS spectra taken at several absorption edges without any additional assumptions about the size of atoms, their actual charges, the presence of magnetic moments and their coupling – all these parameters are known to significantly influence the results of calculations, and, thus, have to be considered with high accuracy. Meanwhile, PDFs calculated from coordinates of relaxed atoms highlight some of the averaged peculiarities of the nearest-neighbor distributions around a particular absorber (the average distances and the mean square displacements) and allow one to correlate, for example, local distortions with magnetic properties since even tiny changes in distances between atoms drastically influence the value and the direction of their magnetic moments for a particular local configuration. Some examples of such an influence depending on HEAs structural arrangement and stoichiometry were demonstrated in Ref. [49] by *ab initio* calculations.

However, due to the similarities of scattering amplitudes for 3d components of the Cantor alloy, the discussions about the distribution of individual components over the coordination shells of a particular absorber have to be skipped. Strong conclusions about a possible short-range ordering could not be done based on the obtained final PDFs. It might only be possible if principal components of the alloy have a significantly different number of valence electrons resulting in easily distinguishable scattering amplitudes or if a particular type of absorber is located in a significantly different local crystallographic surrounding in a multi-phase system. In the latter case, strong differences in partial PDFs should be visible, which is not the case in the present study.

*3.2.1 Local configurations and PDFs*

The starting and final configurations of atoms used in the RMC simulations for the equiatomic CrMnFeCoNi alloy are presented in Fig. 3. The longest interatomic distance potentially accessible in the RMC simulations is equal to half the size of the supercell (~7.1 Å). This value is slightly larger compared to the value used during the RMC fitting (6.0 Å) and was selected according to the position of the minimum between structural peaks in FTs. The particular structural model was simultaneously fitted to EXAFS spectra measured at four absorption edges. The EXAFS calculations were performed within the multiple-scattering formalism considering the scattering paths up to the fourth-order since the higher-order scatterings contribute beyond the range of our analysis (up to 6 Å). This approach has allowed us to perform a reliable analysis of the available experimental EXAFS data and to reconstruct the local environment taking into account the thermal and structural (radial and angular) disorders in the first five coordination shells around each of the 3d absorbers (distances up to ~6.0 Å). To improve statistics and to achieve the required precision of fitting results, a set of twelve independent starting configurations was considered, through this is still many orders of magnitude fewer than the total number of different possible configurations.

The experimental and RMC-fitted EXAFS $\chi(k)k^2$ spectra together with their Fourier and wavelet transforms for the HEA sample are shown in Fig. 4. The agreement between the experiment and calculations is good in both *k*- and *R*-spaces at all considered K edges, and the final structural model represents well all peculiarities in FTs. In particular, we expect that the final structural configuration, which is close to random and is self-consistent, reproduces the arrangement of nearest atoms in the first coordination shell of the corresponding 3d absorbers. This arrangement is responsible for the first peak in FTs in the range from 1.0 to 3.0 Å. An uncertainty in distinguishing between some of the neighboring 3d elements located in the first coordination shell of absorbers of each type does not influence the averaged values of interatomic distances and structural relaxations that will be extracted further.

A set of total and partial PDFs calculated from the atomic coordinates of the final structural model for the HEA is shown in Fig. 5. All constituents exhibit very similar total PDFs averaged over different 3d neighbor atoms (Fig. 5a) with the same number and positions of peaks within the first five coordination shells: this indicates the similarity of their environment in the alloy with a distribution of all atoms over the fcc lattice sites of HEA (within a sensitivity of the applied RMC method). Compared to PDFs of pure metallic foils (see next section), it becomes clear that independently from the initial crystallographic ordering characteristic for pure metals (fcc, bcc or α-phase), all atoms reproduce the fcc structure of the studied Cantor alloy on the local scale.

Nevertheless, there is a well-pronounced splitting of the peak at ~3.5 Å in the total PDF of Fe atoms which is also partly visible for some other elements. Such peak splitting is not observed in pure fcc foils of Co and Ni, and, thus, is specific for the considered sample. It could be due to local structure relaxation (off-center displacement) caused by an existing chemical disorder (atomic heterogeneity) that results in two groups of shorter and longer interatomic distances around Fe atoms. Another explanation could be related to a dynamic effect (double potential), which is not detectable by X-ray diffraction. Considering the partial PDFs (Fig. 5b), it could be noted that only PDFs involving Fe atoms demonstrate such a feature. Thus, we relate it to a specific property of the internal structure of the considered HEA sample, and it may not be observed in other Cantor alloys prepared by different processing routes.

The mean interatomic distances $r$ and the mean square relative displacements (MSRD) $\sigma^2$ were numerically obtained for atoms of each type as the first and second moments of PDFs (see Table 1). In addition, the mean square displacements (MSD) were calculated directly from the final atomic coordinates of atoms in the simulation box relative to their ideal starting positions in the fcc lattice: such information is not typical for EXAFS data analysis and is usually extracted from diffraction data. As could be seen, the mean distances to the neighboring atoms calculated for absorbers of each type are very similar to each other without showing any preferences. Moreover, compared to distances calculated for the hcp/fcc Co and fcc Ni foils, the Cantor alloy demonstrates noticeably larger values of the interatomic distances (compare Tables 1 and 3). This reflects one of the important aspects of atomic heterogeneity in HEAs: Cr and Fe atoms have larger sizes and hence force the atomic volume of the alloy to expand. Such changes, in turn, provoke variations in the interatomic interactions and, thus, in many macroscopic and local properties of this and similar HEAs. It is worth mentioning that pure Fe, Co and Ni are ferromagnetic at room temperature while the Cantor alloy presents a paramagnetic state, this may also be related to the observed difference in interatomic distances.

Besides, it was found quantitatively that Cr atoms have the largest MSRD and MSD values among all constituents of the studied Cantor alloy; and this trend was observed by considering either total or partial PDFs (see Tables 1 and 2, respectively). Earlier, the same conclusion was established for the Al-deficient single-phase fcc $Al_{0.3}CrFeCoNi$ alloy in Ref. [59]. This suggests that the distribution of distances around Cr atoms is larger compared to other constituents, and it may be a reason for distinctive magnetic properties found experimentally in Ref. [35] and shown later in the current work as well.

### 3.3 EXAFS and RMC of pure metallic foils

To demonstrate the accuracy of the RMC method in distinguishing and calculating the relevant parameters for the bcc and fcc structures formed by the same 3d elements as in the studied HEA, EXAFS spectra collected from Cr, Fe, Co and Ni metallic foils shown in Fig. 2b were analyzed. According to Fig. 6, all spectra can be well fitted using the RMC method. The calculated PDFs $g(r)$ are shown in Fig. 7 and demonstrate a clear difference between two crystallographic structures as seen from the peak positions and their relative intensities. At 300 K, the differences between PDFs for the bcc Cr and Fe as well as for the fcc Ni and Co are small in the range up to 4.0 Å, and, therefore, the temperature-dependent measurements are required for better visibility of some peculiarities in their local structure ordering.

The K-edge EXAFS spectrum of the Co foil was additionally fitted by the RMC method considering hcp Co to estimate the deviations in the interatomic distances from Co absorbers to their first coordination shell in fcc and hcp structures. The experimental and fitted EXAFS spectra, their FTs and PDFs for both Co structures (fcc and hcp) are compared in Fig. 8. The PDFs for both Co models are found to be within an error bar for the several first shells illustrating that EXAFS spectroscopy cannot reliably distinguish these two phases from just one set of data at RT. A slightly larger difference in FT around 4.5 Å for the hcp Co model compared to the fcc one suggests that the fcc model is slightly better; however, the difference is too small to prove the existence of either phase. The distant shells, where PDFs demonstrate a clear difference, most probably, could be better resolved at low temperatures providing an opportunity to distinguish these phases from additional measurements. EXAFS data collected within a larger energy range could help to clarify this question as well.

The RMC method applied to the Mn K-edge EXAFS spectrum from the Mn foil agrees well with the expected cubic α-Mn structure. It exhibits a different PDF $g(r)$ for several first coordination shells around the Mn absorber (Fig. 9) than in the fcc phase (Fig. 5). Due to a different structure, the results for pure Mn foil will not be discussed further.

The structural parameters ($r$ and $\sigma^2$) for pure metallic foils were determined from the obtained PDFs in the same way as for the HEA and are summarized in Table 3. The noticeably longer distances and more than twice larger MSRD values are found for the bcc Cr and Fe foils compared to the fcc Co and Ni ones. This is due to the PDFs of the bcc phase, which have two peaks in the range of 1.0-3.0 Å (Fig. 7) caused by eight atoms in the first coordination shell and six atoms in the second shell which were considered together. Therefore, the calculation of PDF moments was performed over two peaks in the case of the bcc structures instead of over just one isolated peak originating from the twelve nearest-neighboring atoms in the case of the fcc/hcp structured metals. The distances calculated for the fcc/hcp Co phases are close to each other; however, both MSRD and MSD are larger for the fcc model compared to the hcp one. It is important that the distances calculated for the fcc/hcp Co and fcc Ni foils are smaller compared to those found for the Cantor alloy as was already mentioned before.

It is worth mentioning that larger MSRD and MSD values found for Cr atoms in the previous section as compared to other constituents in the HEA are not due to averaging over different structural models: all constituents are adopted in the same single-phase fcc crystallographic structure and thus averaging is going over the same number of nearest-neighbors.

**3.4 Magnetometry evaluation**

As was mentioned above, the magnetic properties of HEAs are strongly dependent on the crystallographic structure, composition, and local distortions. To illustrate the complexity of magnetic ordering in the studied HEA flakes at low temperature, the field dependence of the magnetic moment was recorded at 5 K within a small field range of ± 0.5 T after a zero-field cooling (ZFC) and a field-assisted cooling (FC) in the fields of ± 5 T. The resulting field dependencies are shown in Fig. 10. All curves demonstrate a vanishing opening of the hysteresis loop which could be mostly assigned to the remanence of the superconducting coils exploited in the MPMS measurement system [74] rather than to the sample itself. Besides, the hysteresis loops recorded right after FC are shifted in the magnetization direction (vertical shift) up to ± 0.06 emu/g as compared to the symmetric ZFC field-dependence (Fig. 10a). The direction of these vertical shifts is governed by the direction of the applied magnetic field during field- assisted cooling.

The observed shift of the hysteresis loops is much smaller than the one reported in Ref. [35] for the bulk slab of equiatomic Cantor alloy that was several mm thick. However, it clearly demonstrates the presence of frozen magnetic contributions enforced by the direction of FC. These contributions are not fully reversed by the small external magnetic fields of ± 0.5 T and could be associated with the disordered regions around Cr atoms where a strong competition between ferromagnetic and antiferromagnetic couplings of nearest-neighbors is expected. Since all individual PDFs involving Cr atoms demonstrate an increased structural displacement of Cr (Table 2), a fraction of regions with such competing couplings leading to frustrated magnetism could be very large. Taking into account the high heterogeneity of the studied HEA due to the five principal components, these regions are expected to be distributed along the whole volume of the flakes.

This result agrees with previous theoretical findings in Refs. [49] and [35] where individual magnetic moments, as well as a distribution of ferromagnetic and antiferromagnetic interactions, were found to be very sensitive not only to the size of the fcc unit cell but also to the local configurations of the nearest neighbors. It was also shown that Cr atoms prefer to align antiferromagnetically to the cumulative magnetic moment of their first coordination shell regardless of the initial local arrangements [35] and, in general, the antiferromagnetic alignment of Cr with respect to ferromagnetically coupled Fe, Co and Ni is energetically favorable [49]. The same alignment was theoretically predicted for the four-component fcc CrCoFeNi [75] and found experimentally in the sub-surface volume of the $Al_{0.3}$CrCoFeNi HEA [59].

In the case of performing additional field-dependent measurements in the broad field range of ± 5 T right after the field-assisted cooling, the hysteresis loops recorded afterward in the narrow field range do not show any shifts in vertical directions and demonstrate the symmetric behavior as in the case of ZFC (Fig. 10b). This supports the assumption about the presence of initially frustrated or non-collinear magnetic states which can be reduced by applying higher magnetic fields. Further experimental clarifications of relative magnetic coupling of each constituent could be explicitly made only by the synchrotron-based element-specific X-ray magnetic circular dichroism (XMCD) technique and are beyond the scope of the current work.

## 4. Conclusions

The possibility to probe local ordering and inner relaxations in complex multicomponent systems at the atomic scale in an element-specific way opens a path to exhaustive structural studies of materials needed for modern technologies and applications. In this regard, X-ray absorption spectroscopy in conjunction with the reverse Monte Carlo method offers opportunities unattainable by other techniques commonly used to investigate multicomponent alloys. Still, only approximate averaged atomic distributions can be provided without fully probing the details of the enormous number of different actual local atomic configurations in such multicomponent systems. EXAFS spectra recorded at the K absorption edges of several principal components of the alloy can be fitted simultaneously in an unbiased manner using one structural model to reveal the pair distribution functions of the individual elements. Accordingly, quantitative information regarding the number of the nearest neighbors, their distances to the absorber, structural/thermal relaxations of each constituent and peculiarities of the local coordination for each element can be extracted. This information can be further used to infer how individual atoms fit into the crystallographic structure of the complex system which in turn would underline the role of individual components in certain macroscopic properties. Some natural limitations of this approach exist in the case when the multicomponent system consists of elements with very close scattering amplitudes as for the neighboring elements in the periodic table; nevertheless, the most important information regarding absorbers of each type can be still extracted.

Our work demonstrates that all five components of the equiatomic CrMnFeCoNi Cantor alloy are distributed at the nodes of the fcc crystallographic structure independently of their initial ordering in the pure state, a presence of component- specific structural relaxations, and that the alloy is single-phase at the atomic scale. Quantitative analysis reveals statistically averaged interatomic distances for all types of constituents, and as expected these distances are enlarged compared to the distances in pure hcp/fcc foils of Co and Ni. Meanwhile, a much greater disorder was found around Cr atoms suggesting their strong involvement in the formation of magnetically competing couplings over the entire volume of the alloy, which are responsible for the magnetic behavior observed by conventional magnetometry. Surface oxidation probed by XANES reveals the general trend anticipated from previously published results.

Inner relaxations uncovered by element-specific structural findings in this work support the outcomes of previous studies where stronger lattice distortions were associated, in particular, with the large atomic size of Cr atoms while magnetic characterization indicates a strong sensitivity of macroscopic magnetic properties to the synthesis approach. Thus, further activity in developing functional multicomponent materials possessing a high atomic heterogeneity and their detailed studies at the atomic scale will expand our understanding of individual elements relationship with properties of high-entropy systems beyond the limits of the classical Cantor alloy. Besides, further efforts in analysis and interpretation of available experimental data as well as extensive theory modelling are needed to advance the field.

**Acknowledgements**

The authors thank the Helmholtz-Zentrum Berlin for the provision of access to synchrotron radiation facilities and allocation of synchrotron radiation at the BAMline and UE52 beamlines of BESSY II at HZB. The use of ALICE chamber (BMBF project no. 05K19W06) and time for magnetometry measurements at the HZB CoreLab for Quantum Materials is acknowledged as well. Dirk Schröpfer and the workshop from BAM are acknowledged for flakes preparation by mechanical milling; Christiane Stephan-Scherb is acknowledged for providing XRD data. A. Smekhova acknowledges also personal funding from CALIPSOplus project (the Grant Agreement no. 730872 from the EU Framework Programme for Research and Innovation HORIZON 2020). Institute of Solid State Physics, University of Latvia as the Center of Excellence has received funding from the European Union's Horizon 2020 Framework Programme H2020-WIDESPREAD-01-2016-2017-TeamingPhase2 under grant agreement No. 739508, project CAMART2. G. Laplanche acknowledges the German Research Foundation (Deutsche Forschungsgemeinschaft: DFG) for financial support through project LA 3607/3-2 of the Priority Program SPP 2006 "Compositionally Complex Alloys - High Entropy Alloys".

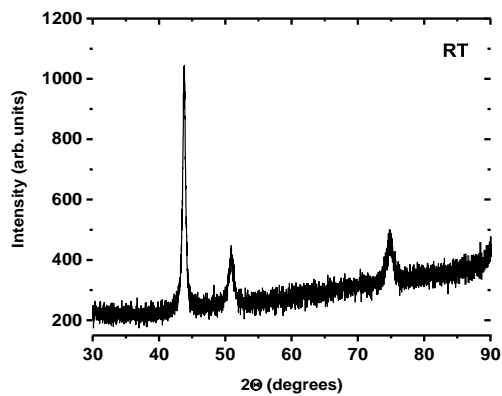

**Fig. 1** XRD patterns from the CrMnFeCoNi alloy recorded at RT.

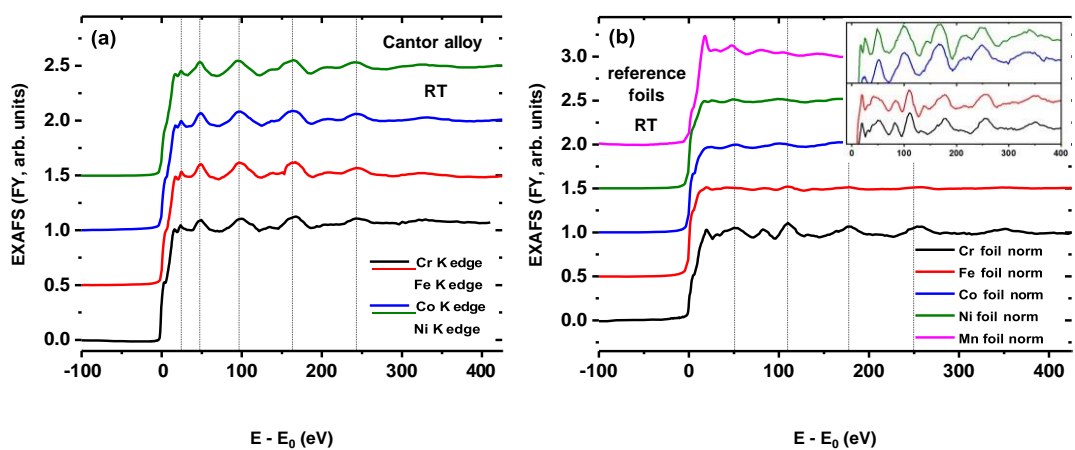

**Fig. 2** X-ray absorption spectra of the single-phase fcc CrMnFeCoNi HEA (a) together with the reference foils (b) recorded at the K absorption edges of Cr ($E_0$ = 5987 eV), Fe ($E_0$ = 7108 eV), Co ($E_0$ = 7708.5 eV), and Ni ($E_0$ = 8333 eV) at RT by FY. The inset shows enlarged EXAFS oscillations to underline the difference between the bcc (Cr and Fe) and fcc (Co and Ni) lattices. EXAFS at the Mn K-edge was measured in transmission ($E_0$ = 6538 eV). $E_0$ was determined as the energy corresponding to the first maximum of the first derivative of each particular spectrum. The spectra are normalized to unity and shifted vertically for clarity.

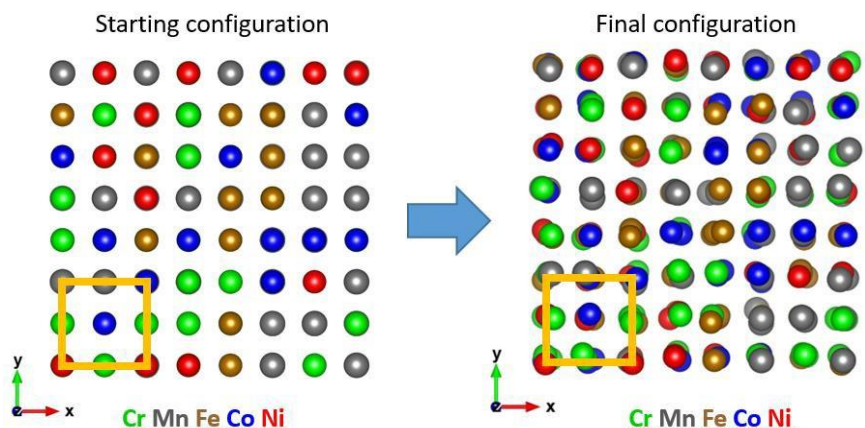

**Fig. 3** Starting and final atom configurations (supercells) used in the RMC simulations for simultaneous fit to EXAFS spectra of the CrMnFeCoNi HEA at four absorption edges. The supercell was randomly filled with constituent atoms according to the stoichiometry of the Cantor alloy; periodic boundary conditions were applied. Color scheme: Cr (green), Mn (gray), Fe (brown), Co (blue), Ni (red). The top views of the supercells are shown along the *z*-direction in chosen *xyz* coordinates as depicted on the figures. The schematic border of the fcc unit cells is displayed at the bottom left of each figure.

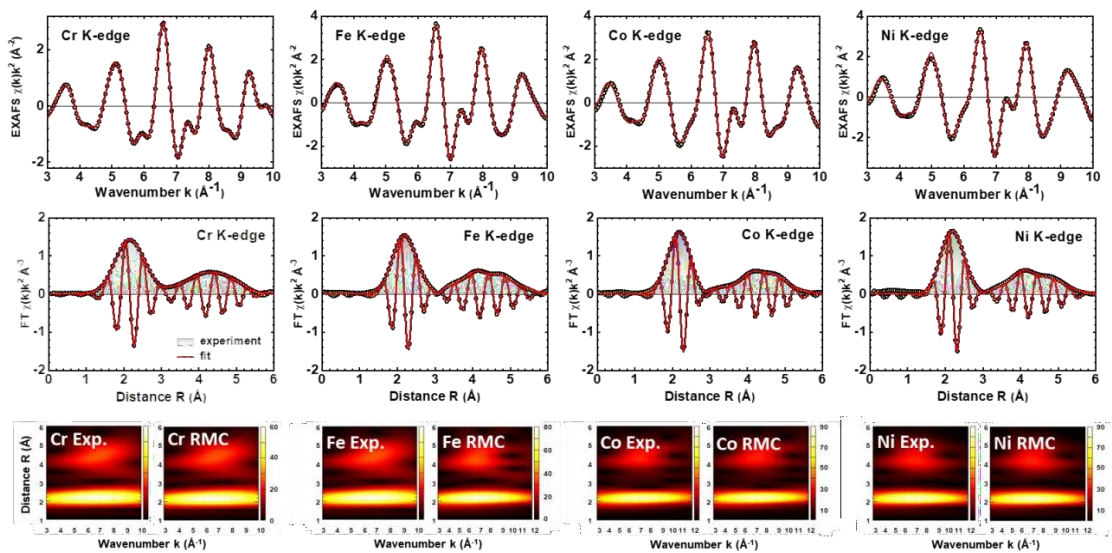

**Fig. 4** Experimental and RMC-calculated Cr, Fe, Co, and Ni K-edge EXAFS spectra $\chi(k)k^2$ and their Fourier and Morlet wavelet transforms for the fcc CrMnFeCoNi HEA at 300 K.

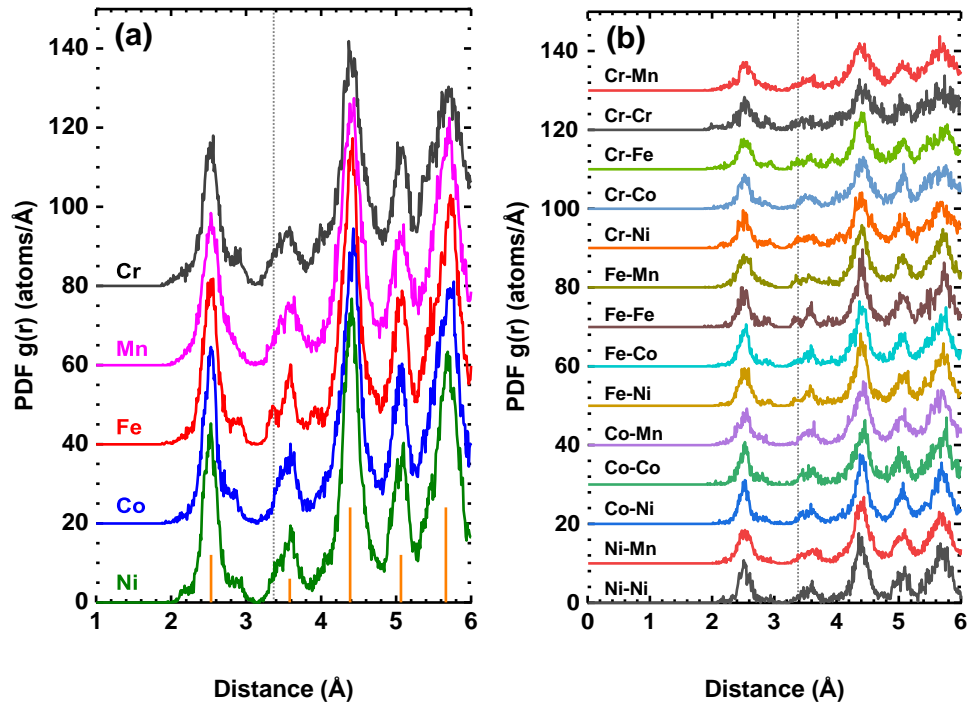

**Fig. 5** Pair distribution functions $g(r)$ – (a) total and (b) partial – for the fcc CrMnFeCoNi HEA at 300 K extracted from K-edge EXAFS spectra of Cr, Fe, Co, and Ni using the RMC method.

**Table 1** Structural parameters extracted from the total PDFs of the fcc CrMnFeCoNi HEA.

|    | Phase in pure metal | $r$ [Å] (±0.02 Å) | MSRD $\sigma^2$ [Å$^2$] (±0.003 Å$^2$) | MSD [Å] (±0.02 Å) |
|---|---|---|---|---|
| **Cr** | bcc | 2.55 | **0.039** | **0.24** |
| **Mn** | α-Mn | 2.55 | 0.028 | 0.18 |
| **Fe** | bcc | 2.54 | 0.028 | 0.17 |
| **Co** | hcp/fcc | 2.54 | 0.028 | 0.19 |
| **Ni** | fcc | 2.54 | 0.027 | 0.19 |

(All average distances are about 2.54-2.55 Å; Cr atoms have larger MSRD and MSD)

**Table 2** Structural parameters extracted from the partial PDFs of the studied Cantor alloy.

| $r$ [Å] / MSRD [Å$^2$] | Cr | Mn | Fe | Co | Ni |
|---|---|---|---|---|---|
| Cr | 2.55 / **0.050** | 2.55 / **0.038** | 2.55 / **0.038** | 2.54 / **0.033** | 2.55 / **0.037** |
| Mn |  | 2.55 / 0.026 | 2.54 / 0.026 | 2.55 / 0.028 | 2.54 / 0.025 |
| Fe |  |  | 2.55 / 0.028 | 2.54 / 0.026 | 2.54 / 0.023 |
| Co |  |  |  | 2.54 / 0.025 | 2.54 / 0.025 |
| Ni |  |  |  |  | 2.54 / 0.023 |

(All average distances are about 2.54-2.55 Å; Cr atoms have the largest MSRD)

**Table 3** Structural parameters extracted from the PDFs of pure metallic foils.

|  | phase | $r$ [Å] (±0.02 Å) | N | MSRD $\sigma^2$ [Å$^2$] (±0.003 Å$^2$) | MSD [Å] | MSD $\sigma^2$ [Å$^2$] |
|---|---|---|---|---|---|---|
| **Cr** | bcc | 2.67 | 8 + 6 | 0.048 | 0.13 | 0.017 |
| **Fe** | bcc | 2.65 | 8 + 6 | 0.045 | 0.12 | 0.014 |
| **Co** | fcc | 2.52 | 12 | 0.020 | 0.15 | 0.023 |
| **Co** | hcp | 2.51 | 12 | 0.016 | 0.13 | 0.017 |
| **Ni** | fcc | 2.50 | 12 | 0.016 | 0.13 | 0.017 |

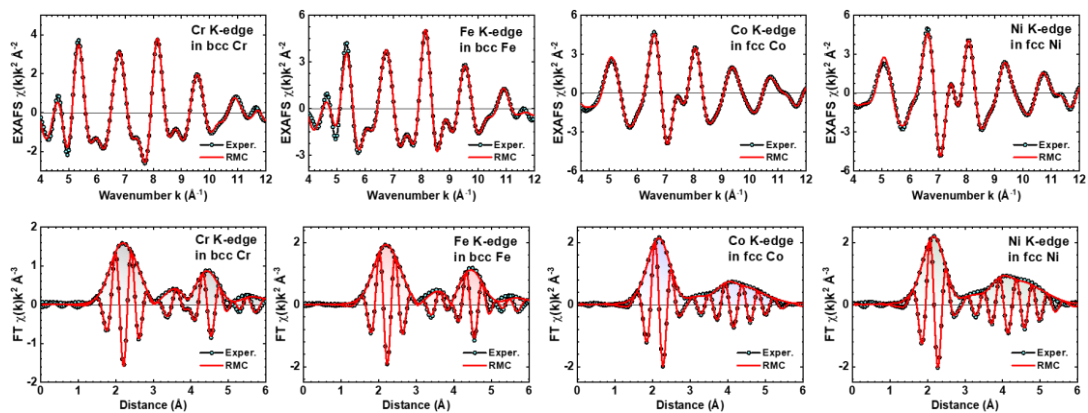

**Fig. 6** Experimental and RMC-calculated Cr, Fe, Co, and Ni K-edge EXAFS spectra $\chi(k)k^2$ of the pure metallic foils and their Fourier transforms at 300 K.

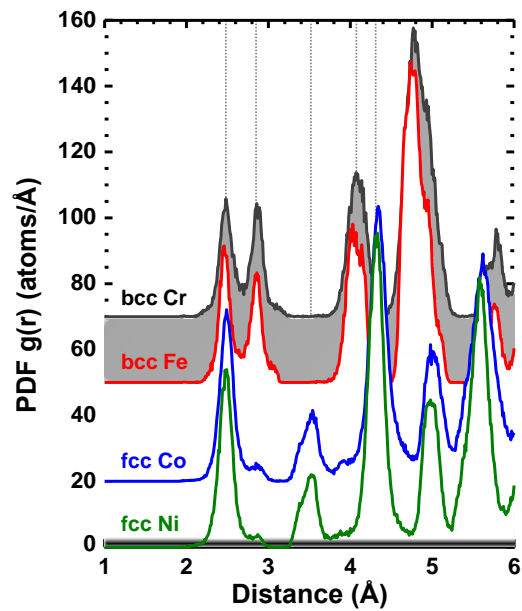

**Fig. 7** Pair distribution functions $g(r)$ for the pure metallic foils at 300 K extracted from K-edge EXAFS spectra of Cr, Fe, Co, and Ni using the RMC method.

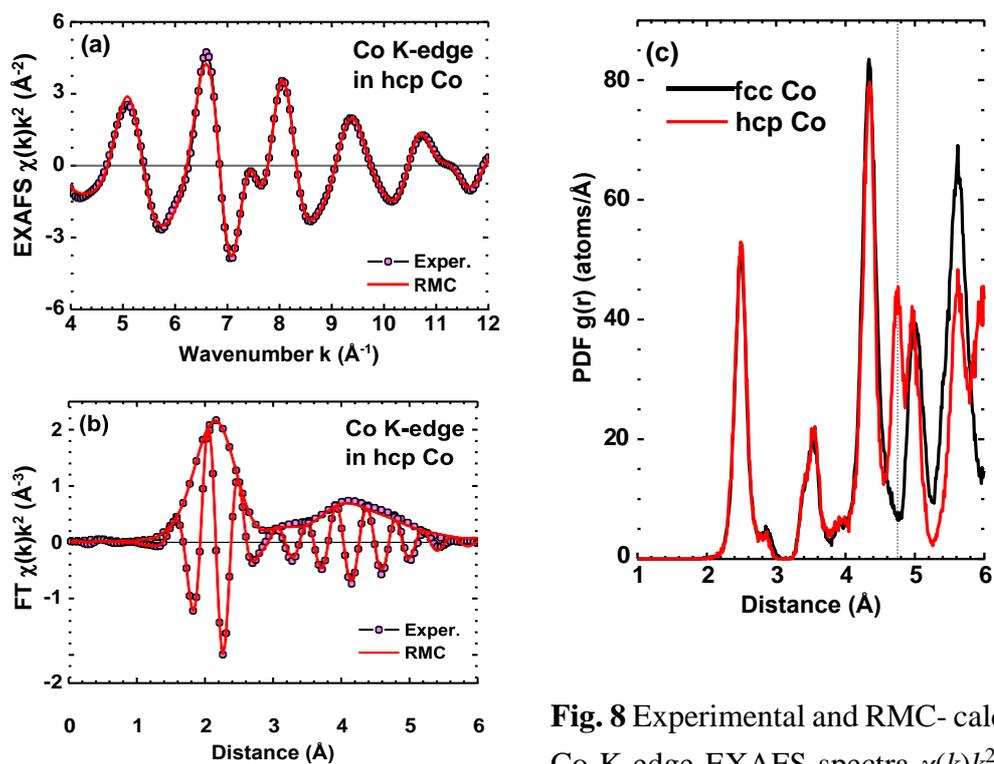

**Fig. 8** Experimental and RMC- calculated Co K-edge EXAFS spectra $\chi(k)k^2$ of the CrMnFeCoNi HEA (a) and their Fourier transforms (b) for the model with hcp Co structure at 300 K.

(c) The comparison of PDFs calculated for the fcc and hcp Co models.

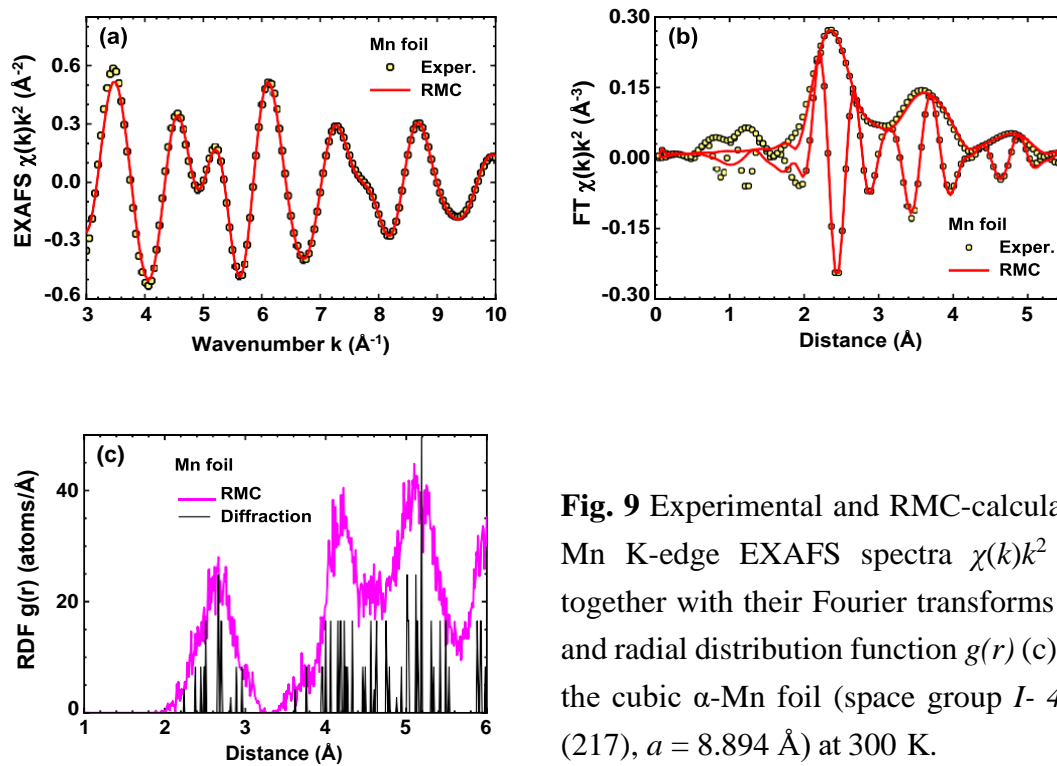

**Fig. 9** Experimental and RMC-calculated Mn K-edge EXAFS spectra $\chi(k)k^2$ (a) together with their Fourier transforms (b) and radial distribution function $g(r)$ (c) for the cubic α-Mn foil (space group $I\text{-}43m$ (217), $a$ = 8.894 Å) at 300 K.

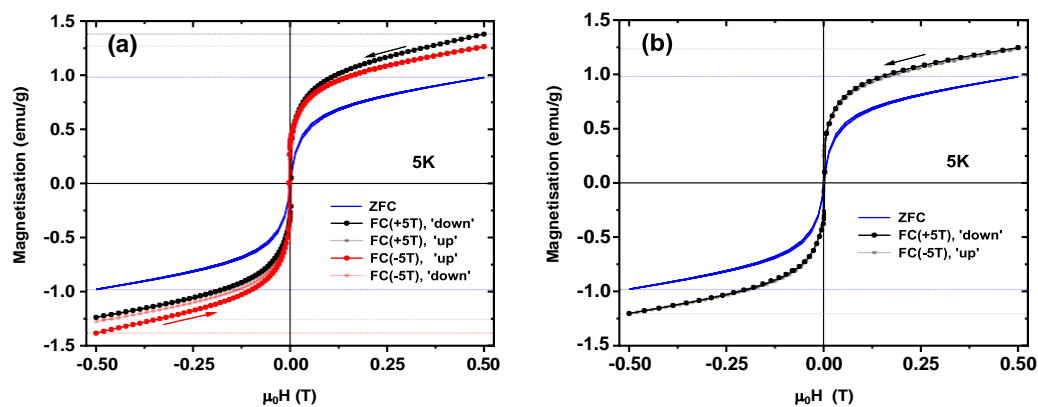

**Fig. 10** Field-dependences of the HEA flakes recorded at 5 K after the zero-field cooling and field-assisted cooling in ± 5 T performed right after the cooling process (a) and after an additional field-dependence in a broad field range of ± 5 T (b). The vertical shift of the hysteresis loops is observed in (a).

**Appendix A1 Surface oxidation probed by XANES**

XANES spectra recorded with soft X-rays at the $L_{2,3}$ absorption edges of all 3d constituents were used to check oxidation states of ions in the sub-surface (down to several nm) volume of the CrMnFeCoNi HEA flakes (Fig. A1). A degree of surface oxidation for Cr, Fe, Co and Ni constituents was found to be very similar to those reported previously for the fcc $Al_{0.3}$CrFeCoNi alloy using XANES [59] and CrFeCoNi using X-ray photoelectron spectroscopy (XPS) [76, 77].

The lineshape of XANES spectra recorded at the Cr $L_{2,3}$ absorption edges indicates that Cr forms $Cr_2O_3$ oxide and is in the $Cr^{3+}$ oxidation state. Mn atoms are supposed to be oxidized down to Mn monoxide since the lineshape of Mn XANES spectrum is in-between of MnO/MnS reference spectra [78, 79] suggesting the presence of $Mn^{2+}$ ions in the top oxide layer. Fe and Co atoms can form different types of oxides at the surface of HEAs, and here Fe atoms show the $Fe^{3+}$-like ion state while Co is only slightly oxidized. As expected, Ni atoms nearly do not participate in the formation of the passivation layer and are in the metallic-like state.

Corrosion, oxidation and catalytic activity of HEAs were considered in detail earlier [4, 80]. There were also attempts to estimate the migration energy barriers and to observe experimentally the diffusivity of individual atoms to understand the main trends in the surface activity. For the Cantor alloy these studies revealed slightly different relations between energy barriers (as Ni > Co > Cr > Fe > Mn [81]) and diffusivity (as Ni < Co < Fe < Cr < Mn [5]). A common assumption about a number of 3d holes sequence for pure 3d elements provides another series as Cr > Mn > Fe > Co > Ni. Thus, it was intriguing to check the formed oxides at the top of a particular HEA sample by X-ray absorption spectroscopy. Further studies would require depth-resolved methods to better understand the most favorable conditions at which the oxidation, catalytic, heat and corrosion performance of equiatomic or off-stoichiometry HEAs (and the nanostructures based on them) could replace common steels and existing composite materials in various vital applications.

https://doi.org/10.1016/j.jallcom.2022.165999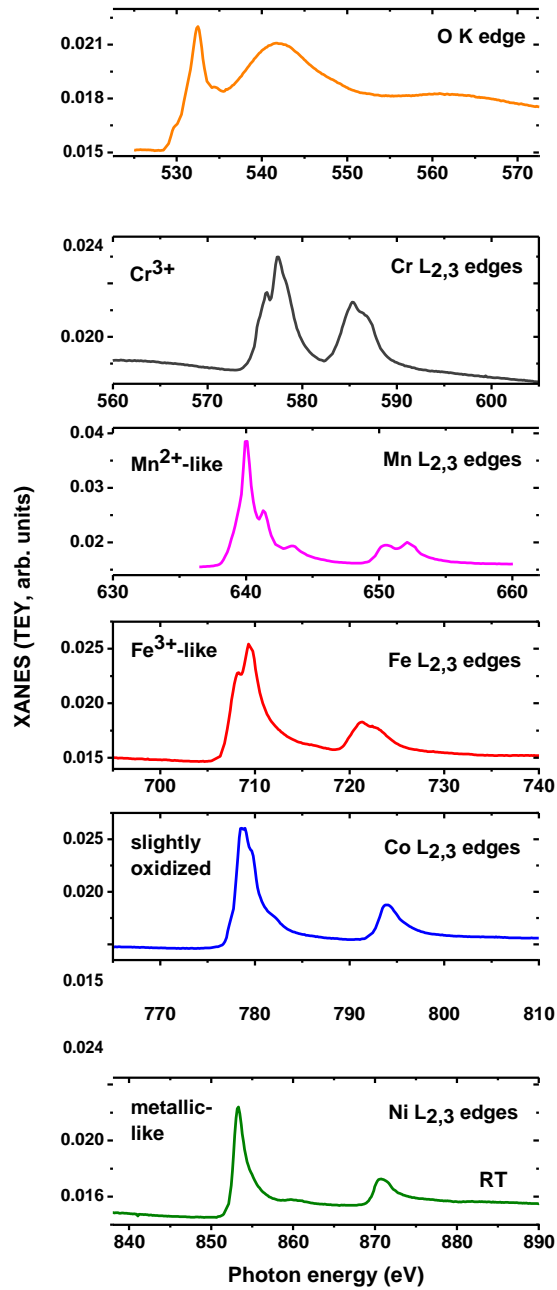

**Fig. A1** Raw XANES spectra recorded from the fcc CrMnFeCoNi HEA with horizontally polarized X-rays at RT by TEY at the $L_{2,3}$ absorption edges of Cr, Mn, Fe, Co, and Ni, and at the K edge of oxygen.

# References

[1] B. Cantor, I. T. H. Chang, P. Knight, A. J. B. Vincent, Microstructural development in equiatomic multicomponent alloys, Mater. Sci. Eng. 375–377 (2004) 213–218.

[2] J. W. Yeh, S. K. Chen, S. J. Lin, J. Y. Gan, T. S. Chin, T. T. Shun, C.-H. Tsau, S.-Y. Chang, Nanostructured high-entropy alloys with multiple principal elements: novel alloy design concepts and outcomes, Adv. Eng. Mater. 6 (2004) 299–303.

[3] D. B. Miracle, O. N. Senkov, A critical review of high entropy alloys and related concepts, Acta Mater. 122 (2017) 448–511.

[4] Y. Ma, Y. Ma, Q. Wang, S. Schweidler, M. Botros, T. Fu, H. Hahn, T. Brezesinski, B. Breitung. High-entropy energy materials: challenges and new opportunities, Energy Environ. Sci. 14 (2021) 2883–2905.

[5] K.Y. Tsai, M.H. Tsai, J.W. Yeh, Sluggish diffusion in Co–Cr–Fe–Mn–Ni high entropy alloys, Acta Mater. 61 (2013) 4887–4897.

[6] F. Otto, Y. Yang, H. Bei, E. P. George, Relative effects of enthalpy and entropy on the phase stability of equiatomic high-entropy alloys. Acta Mater. 61 (2013) 2628–2638.

[7] A. Durand, L. Peng, G. Laplanche, J.R. Morris, E.P. George, G. Eggeler, Interdiffusion in Cr–Fe–Co–Ni medium-entropy alloys, Intermetallics 122 (2020) 106789.

[8] B. Gludovatz, A. Hohenwarter, D. Catoor, E.H. Chang, E.P. George, R.O. Ritchie, A fracture-resistant high-entropy alloy for cryogenic applications, Science 345 (2014) 1153.

[9] Y.-F. Kao, S.-K. Chen, J.-H. Sheu, J.-T. Lin, W.-E. Lin, J.-W. Yeh, S.-J. Lin, T.-H. Liou, Ch.-W. Wang, Hydrogen storage properties of multi-principal-component CoFeMnTi$_x$V$_y$Zr$_z$ alloys, Intern. J. Hydrogen Ene. 35 (2010) 9046–9059.

[10] M. Sahlberg, D. Karlsson, C. Zlotea, U. Jansson, Superior hydrogen storage in high entropy alloys, Sci. Rep. 6 (2016) 36770.

[11] T. Löffler, H. Meyer, A. Savan, P. Wilde, A. Garzón Manjón, Y.-T. Chen, E. Ventosa, Ch. Scheu, A. Ludwig, W. Schuhmann, Discovery of a multinary noble metal–free oxygen reduction catalyst, Adv. En. Mat. 8 (2018) 1802269.

[12] G. Fang, J. Gao, J. Lv, H. Jia, H. Li, W. Liu, G. Xie, Z. Chen, Y. Huang, Q. Yuan, X. Liu, X. Lin, Sh. Sun, H.-J. Qiu, Multi-component nanoporous alloy/(oxy)hydroxide for bifunctional oxygen electrocatalysis and rechargeable Zn-air batteries, Appl. Cat. B: Environ. 268 (2020) 118431.

[13] T. Löffler, A. Savan, H. Meyer, M. Meischein, V. Strotkötter, A. Ludwig, W. Schuhmann, Design of complex solid-solution electrocatalysts by correlating configuration, adsorption energy distribution patterns, and activity curves, Angew. Chem. Int. Ed. 59 (2020) 5844–5850.

[14] Y. Yao, Zh. Huang, T. Li, H. Wang, Y. Liu, H. S. Stein, Y. Mao, J. Gao, M. Jiao, Q. Dong, J. Dai, P. Xie, H. Xie, S. D. Lacey, I. Takeuchi, J. M. Gregoire, R. Jiang, Ch. Wang, A. D. Taylor, R. Shahbazian-Yassar, L. Hu, High-throughput, combinatorial synthesis of multimetallic nanoclusters, PNAS 117 (2020) 6316–6322.

[15] J. K. Pedersen, Th. A. A. Batchelor, A. Bagger, J. Rossmeisl, High-entropy alloys as catalysts for the $CO_2$ and CO reduction reactions, ACS Catal. 10 (2020) 2169–2176.

[16] K. Kong, J. Hyun, Y. Kim, W. Kim, D. Kim, Nanoporous structure synthesized by selective phase dissolution of AlCoCrFeNi high entropy alloy and its electrochemical properties as supercapacitor electrode, J. Power Sourc. 437 (2019) 226927.

[17] X. Xu, Y. Du, Ch. Wang, Y. Guo, J. Zou, K. Zhou, Zh. Zeng, Y. Liu, L. Li, High-entropy alloy nanoparticles on aligned electronspun carbon nanofibers for supercapacitors, J. Alloy. Comp. 822 (2020) 153642.

[18] Y. Zhang, Sh. Zhao, W. J. Weber, K. Nordlund, F. Granberg, F. Djurabekova, Atomic-level heterogeneity and defect dynamics in concentrated solid-solution alloys, Curr. Opin. Solid State Mater. Sci. 21 (2017) 221–237.

[19] E. J. Pickering, A. W. Carruthers, P. J. Barron, S. C. Middleburgh, D. E. J. Armstrong, A. S. Gandy, High-entropy alloys for advanced nuclear applications, Entropy 23 (2021) 98.

[20] R. Zhang, S. Zhao, J. Ding, Y. Chong, T. Jia, C. Ophus, M. Asta, R. O. Ritchie, A. M. Minor, Short-range order and its impact on the CrCoNi medium-entropy alloy, Nature 581 (2020) 283–287.

[21] K. A. Christofidou, T. P. McAuliffe, P. M. Mignanelli, H. J. Stone, N. G. Jones, On the prediction and the formation of the sigma phase in $CrMnCoFeNi_x$ high entropy alloys, J. Alloys Compd. 770 (2019) 285–293.

[22] M. E. Bloomfield, K. A. Christofidou, N. G. Jones, Effect of Co on the phase stability of $CrMnFeCo_xNi$ high entropy alloys following long-duration exposures at intermediate temperatures, Intermetallics 114 (2019) 106582.

[23] M. E. Bloomfield, K. A. Christofidou, F. Monni, Q. Yang, M. Hang, N. G. Jones, The influence of Fe variations on the phase stability of $CrMnFe_xCoNi$ alloys following long-duration exposures at intermediate temperatures, Intermetallics 131 (2021) 107108.

[24] B. Schuh, F. Mendez-Martin, B. Völker, E. P. George, H. Clemens, R. Pippan, A. Hohenwarter, Mechanical properties, microstructure and thermal stability of a nanocrystalline CoCrFeMnNi high-entropy alloy after severe plastic deformation, Acta Mater. 96 (2015) 258–268.

[25] G. Laplanche, S. Berglund, C. Reinhart, A. Kostka, F. Fox, E.P. George, Phase stability and kinetics of σ-phase precipitation in CrMnFeCoNi high-entropy alloys, Acta Mater. 161 (2018) 338–351.

[26] C. Li, J. C. Li, M. Zhao, Q. Jiang, Effect of aluminum contents on microstructure and properties of $Al_xCoCrFeNi$ alloys, J. Alloys Compd. 504 (2010) S515–S518.

[27] Y.-F. Kao, T.-J. Chen, S.-K. Chen, J.-W. Yeh, Microstructure and mechanical property of as-cast, -homogenized, and -deformed $Al_xCoCrFeNi$ ($0 \leq x \leq 2$) high-entropy alloys, J. Alloys Compd. 488 (2009) 57–64.

[28] J. Cieslak, J. Tobola, M. Reissne, The effect of bcc and fcc phase preference on magnetic properties of $Al_xCrFeCoNi$ high entropy alloys, Intermetallics 118 (2020) 106672.

[29] K. Yusenko, S. Riva, W. Crichton, K. Spektor, E. Bykova, A. Pakhomova, A. Tudball, I. Kupenko, A. Rohrbach, S. Klemme, F. Mazzali, S. Margadonna, N. P. Lavery, S. G. R. Brown, High-pressure high-temperature tailoring of high entropy alloys for extreme environments, J. Alloys Compd. 738 (2018) 491–500.

[30] O. Stryzhyboroda, V. T. Witusiewicz, S. Gein, D. Röhrens, U. Hecht, Phase equilibria in the Al–Co–Cr–Fe–Ni high entropy alloy system: Thermodynamic description and experimental study, Front. in Mater. 7 (2020) 270.

[31] F. Otto, A. Dlouhý, Ch. Somsen, H. Bei, G. Eggeler, E.P. George, The influences of temperature and microstructure on the tensile properties of a CoCrFeMnNi high-entropy alloy, Acta Mater. 61 (2013) 5743–5755.

[32] A. Gali, E. P. George, Tensile properties of high-and medium-entropy alloys, Intermetallics 39 (2013) 74–78.

[33] H. Luo, Z. Li, A. M. Mingers, D. Raabe, Corrosion behavior of an equiatomic CoCrFeMnNi high-entropy alloy compared with 304 stainless steel in sulfuric acid solution, Corros. Sci. 134 (2018) 131–139.

[34] M. Zhu, B. Zhao, Y. Yuan, Sh. Guo, G. Wei, Study on corrosion behavior and mechanism of CoCrFeMnNi HEA interfered by AC current in simulated alkaline soil environment, J. Electr. Chem. 882 (2021) 115026.

[35] O. Schneeweiss, M. Friák, M. Dudová, D. Holec, M. Šob, D. Kriegner, V. Holý, P. Beran, E. P. George, J. Neugebauer, A. Dlouhý, Magnetic properties of the CrMnFeCoNi high-entropy alloy, Phys. Rev. B 96 (2017) 014437.

[36] A. Zaddach, C. Niu, C. Koch, D. Irving, Mechanical properties and stacking fault energies of NiFeCrCoMn high-entropy alloy, JOM 65 (2013) 1780–1789.

[37] Sh. Huang, W. Li, S. Lu, F. Tian, J. Shen, E. Holmström, L. Vitos, Temperature dependent stacking fault energy of FeCrCoNiMn high entropy alloy, Scr. Mater. 108 (2015) 44–47.

[38] C. Varvenne, A. Luque, W. A. Curtin, Theory of strengthening in fcc high entropy alloys, Acta Mater. 118 (2016) 164–176.

[39] B. Wang, A. Fu, X. Huang, B. Liu, Z. Li, X. Zan, Mechanical properties and microstructure of the CoCrFeMnNi high entropy alloy under high strain rate compression, J. Mater. Eng. Perform. 25 (2016) 2985–2992.

[40] G. Bracq, M. Laurent-Brocq, L. Perrière, R. Pirès, J.-M. Joubert, I. Guillot, The fcc solid solution stability in the Co-Cr-Fe-Mn-Ni multi-component system, Acta Mater. 128 (2017) 327–336.

[41] G. Bracq, M. Laurent-Brocq, C. Varvenne, L. Perrière, W. A. Curtin, J.-M. Joubert, I. Guillot, Combining experiments and modelling to explore the solid solution strengthening of high and medium entropy alloys, Acta Mater. 177 (2019) 266–279.

[42] M. Laurent-Brocq, L. Perrière, R. Pirès, Y. Champion, From high entropy alloys to diluted multi-component alloys: range of existence of a solid-solution, Mater. Design 103 (2016) 84–89.

[43] M. Laurent-Brocq, L. Perrière, R. Pirès, F. Prima, P. Vermaut, Y. Champion, From diluted solid solutions to high entropy alloys: on the evolution of properties with composition of multi-component alloys, Materials Science and Engineering A696 (2017) 228–235.

[44] B. Cantor, Multicomponent high-entropy Cantor alloys, Prog. Mater. Science. 120 (2020) 100754.

[45] Y. Tong, G. Velisa, T. Yang, K. Jin, C. Lu, H. Bei, J. Y. P. Ko, D. C. Pagan, R. Huang, Y. Zhang, L. Wang, F. X. Zhang, Probing local lattice distortion in medium- and high-entropy alloys, arXiv [preprint] (2017) arXiv:1707.07745.

[46] Q. Ding, Y. Zhang, X. Chen, X. Fu, D. Chen, S Chen, L. Gu, F. Wei, H. Bei, Y. Gao, M. Wen, J. Li, Z. Zhang, T. Zhu, R. O. Ritchie, Q. Yu, Tuning element distribution, structure and properties by composition in high-entropy alloys, Nature 574 (2019) 223–227.

[47] L. Cižek, P. Kratochvíl, B. Smola, Solid solution hardening of copper crystals, J. Mater. Sci. 9 (1974) 1517–1520.

[48] L. A. Gypen, A. Deruyttere, Multi-component solid solution hardening, J. Mater. Sci. 12 (1977) 1028–1033.

[49] D. Ma, B. Grabowski, F. Körmann, J. Neugebauer, D. Raabe, Ab initio thermodynamics of the CoCrFeMnNi high entropy alloy: Importance of entropy contributions beyond the configurational one, Acta Mater. 100 (2015) 90–97.

[50] N. L. Okamoto, K. Yuge, K. Tanaka, H. Inui, E. P. George, Atomic displacement in the CrMnFeCoNi high-entropy alloy–A scaling factor to predict solid solution strengthening, AIP Advances 6 (2016) 125008.

[51] L. R. Owen, E. J. Pickering, H. Y. Playford, H. J. Stone, M. G. Tucker, N. G. Jones, An assessment of the lattice strain in the CrMnFeCoNi high-entropy alloy, Acta Mater. 122 (2017) 11–18.

[52] H. S. Oh, K. Odbadrakh, Yu. Ikeda, S. Mu, F. Körmann, Ch.-J. Sun, H. S. Ahn, K. N. Yoon, D. Ma, C. C. Tasan, T. Egami, E. S. Park, Element-resolved local lattice distortion in complex concentrated alloys: An observable signature of electronic effects, Acta Mater. 216 (2021) 117135.

[53] F. Zhang, Y. Tong, K. Jin, H. Bei, W. J. Weber, A. Huq, Chemical complexity induced local structural distortion in NiCoFeMnCr high-entropy alloy, Mat. Res. Lett. 6 (2018) 450–455.

[54] H. S. Oh, D. Ma, G. P. Leyson, B. Grabowski, E. S. Park, F. Körmann, D. Raabe, Lattice distortions in the FeCoNiCrMn high entropy alloy studied by theory and experiment, Entropy 18 (2016) 321.

[55] Q. F. He, P. H. Tang, H. A. Chen, S. Lan, J. G. Wang, J. H. Luan, M. Du, Y. Liu, C. T. Liu, C. W. Pao, Y. Yang, Understanding chemical short-range ordering/demixing coupled with lattice distortion in solid solution high entropy alloys, Acta Mater. 216 (2021) 117140.

[56] Q. He, Y. Yang, On lattice distortion in high entropy alloys, Front. in Mater. 5 (2018) 42.

[57] L. R. Owen, N. G. Jones, Lattice distortions in high-entropy alloys, J. Mat. Res. 33 (2018) 2954– 2969.

[58] L. R. Owen, N. G. Jones, Quantifying local lattice distortions in alloys, Scr. Mater. 187 (2020) 428– 433.

[59] A. Smekhova, A. Kuzmin, K. Siemensmeyer, C. Luo, K. Chen, F. Radu, E. Weschke, U. Reinholz, A. Guilherme Buzanich, K. V. Yusenko, Al-driven peculiarities of local coordination and magnetic properties in single-phase Al$_x$-CrFeCoNi high-entropy alloys, Nano Res. (2021), https://doi.org/10.1007/s12274-021-3704-5.

[60] M. Schneider, E. P. George, T. J. Manescau, T. Záležák, J. Hunfeld, A. Dlouhý, G. Eggeler, G. Laplanche, Analysis of strengthening due to grain boundaries and annealing twin boundaries in the CrCoNi medium-entropy alloy, Int. J. Plast. 124 (2020) 155–169.

[61] G. Laplanche, Growth kinetics of σ-phase precipitates and underlying diffusion processes in CrMnFeCoNi high-entropy alloys, Acta Mater. 199 (2020) 193–208.

[62] H. Riesemeier, K. Ecker, W. Goerner, B. R. Muller, M. Radtke, M. Krumrey, Layout and first XRF applications of the BAMline at BESSY II, X-Ray Spectrom. 34 (2005) 160–163.

[63] Ch. Lutz, S. Hampel, X. Ke, S. Beuermann, Th. Turek, U. Kunz, A. Guilherme Buzanich, M. Radtke, U. E. A. Fittschen, Evidence for redox reactions during vanadium crossover inside the nanoscopic water-body of Nafion 117 using X-ray absorption near edge structure spectroscopy, J. Power Sour. 483 (2021) 229176.

[64] A. Kuzmin, J. Chaboy, EXAFS and XANES analysis of oxides at the nanoscale, IUCrJ 1 (2014) 571– 589.

[65] Xaesa, v0.04; GitHub: 2021. https://github.com/aklnk/xaesa (accessed Nov 1, 2020).

[66] J. Timoshenko, A. Kuzmin, J. Purans, Reverse Monte Carlo modeling of thermal disorder in crystalline materials from EXAFS spectra, Comput. Phys. Commun. 183 (2012) 1237–1245.

[67] J. Timoshenko, A. Kuzmin, J. Purans, EXAFS study of hydrogen intercalation into ReO3 using the evolutionary algorithm, J. Phys. Condens. Matter 26 (2014) 055401.

[68] A. L. Ankudinov, B. Ravel, J. J. Rehr, S. D. Conradson, Real space multiple-scattering calculation and interpretation of X-ray absorption near-edge structure, Phys. Rev. B. 58 (1998) 7565–7576.

[69] J. J. Rehr, R. C. Albers, Theoretical approaches to X-ray absorption fine structure, Rev. Mod. Phys. 72 (2000) 621–654.

[70] L. Hedin, B. I. Lundqvist, Explicit local exchange-correlation potentials, J. Phys. C: Solid State Phys. 4 (1971) 2064–2083.

[71] R. Abrudan, F. Brüssing, R. Salikhov, J. Meermann, Radu, H. Ryll, F. Radu, H. Zabel, ALICE – An advanced reflectometer for static and dynamic experiments in magnetism at synchrotron radiation facilities, Rev. Sci. Instr. 86 (2015) 063902.

[72] R. Abrudan, F. Radu, ALICE: A diffractometer/reflectometer for soft X-ray resonant magnetic scattering at BESSY II, Journal of large-scale research facilities 2 (2016) A69.

[73] P. S. Miedema, W. Quevedo, M. Fondell, The variable polarization undulator beamline UE52 SGM at BESSY II, Journal of large-scale research facilities 2 (2016) A70.

[74] The remnant coercive fields after the field-assisted cooling in ±5T magnetic field were found to be up to ±33 Oe according to the same measurement sequence applied to the reference Pd sample. After the field-assisted cooling in +5T (-5T), the observed coercivity was +33 Oe (-32 Oe) for the first field-sweep and -17 Oe (+18 Oe) for the second field-sweep.

[75] C. Niu, A.J. Zaddach, A.A. Oni, X. Sang, J.W. Hurt III, J.M. LeBeau, C. C. Koch, D. L. Irving, Spin-driven ordering of Cr in the equiatomic high entropy alloy NiFeCrCo, Appl. Phys. Lett. 106 (2015) 161906.

[76] C. B. Nascimento, U. Donatus, C. T. Ríos, R. A. Antunes, Electronic properties of the passive films formed on CoCrFeNi and CoCrFeNiAl high entropy alloys in sodium chloride solution, J. Mater. Res. Technol. 9 (2020) 13879–13892.

[77] Y. Z. Shi, B. Yang, P. D. Rack, S. F. Guo, P. K. Liaw, Y. Zhao, High-throughput synthesis and corrosion behavior of sputter deposited nanocrystalline $Al_x(CoCrFeNi)_{100-x}$ combinatorial high entropy alloys, Mater. Design 195 (2020) 109018.

[78] C. Mitra, Z. Hu, P. Raychaudhuri, S. Wirth, S. I. Csiszar, H. H. Hsieh, H.-J. Lin, C. T. Chen, L. H. Tjeng, Direct observation of electron doping in $La_{0.7}Ce_{0.3}MnO_3$ using x-ray absorption spectroscopy, Phys. Rev. B 67 (2003) 092404.

[79] S. P. Cramer, F. M. F. DeGroot, Y. Ma, C. T. Chen, F. Sette, C. A. Kipke, D. M. Eichhorn, M. K. Chan, W. H. Armstrong, E. Libby, G. Christou, S. Brooker, V. McKee, O. C. Mullins, J. C. Fuggle, Ligand field strengths and oxidation states from manganese L-edge spectroscopy, J. Am. Chem. Soc. 113 (1991) 7937–7940.

[80] B. R. Anne, S. Shaik, M. Tanaka, A. Basu, A crucial review on recent updates of oxidation behaviour in high entropy alloys, SN Appl. Sci. 3 (2021) 366.

[81] W. M. Choi, Y. H. Jo, S. S. Sohn, S. Lee, B.-J. Lee, Understanding the physical metallurgy of the CoCrFeMnNi high-entropy alloy: an atomistic simulation study, NPJ Comput. Mater. 4 (2018) 1.

## CRediT authorship contribution statement

**Alevtina Smekhova:** Conceptualization, Investigation (XANES), Formal analysis (Magnetometry), Writing - Original Draft, Review & Editing, Visualization, Supervision; **Alexei Kuzmin:** Formal analysis (EXAFS, RMC), Writing - Review & Editing, Visualization; **Konrad Siemensmeyer:** Investigation (Magnetometry);

**Radu Abrudan:** Investigation (XANES); **Uwe Reinholz:** Investigation (EXAFS); **Ana Guilherme Buzanich:** Investigation (EXAFS); **Mike Schneider:** Resources (Sample preparation); **Guillaume Laplanche:** Resources (Sample preparation), Writing - Review & Editing; **Kirill V. Yusenko:** Conceptualization, Investigation (EXAFS), Writing - Original Draft, Review & Editing, Supervision; All authors read, discussed, commented and approved the final version of the manuscript.

**Declaration of interests**

☒ The authors declare that they have no known competing financial interests or personal relationships that could have appeared to influence the work reported in this paper.

**Highlights**

- Multi-edge EXAFS spectroscopy was used to explore element-specific local ordering
- Reverse Monte Carlo simulations allowed to fit simultaneously several EXAFS spectra
- Pair distribution functions of each individual component were unbiasedly revealed
- Enlarged structural displacements were found solely for Cr atoms
- The observed magnetic behavior was assigned to effects related to Cr relaxations